\documentclass[aps,prl,preprint,superscriptaddress,fleqn,showkeys]{revtex4-2}
\usepackage{subeqnarray}
\usepackage{amsmath,amssymb}
\usepackage{graphicx}
\usepackage{subfigure}
\usepackage{float}
\usepackage{dcolumn}
\usepackage{bm}
\usepackage{subeqnarray}
\usepackage{cases}
\usepackage{hyperref}
\usepackage[usenames]{color}
\def\be{\begin{equation}}
\def\ee{\end{equation}}
\def\bea{\begin{eqnarray}}
\def\eea{\end{eqnarray}}
\def\bs{\begin{subequations}}
\def\es{\end{subequations}}

\begin{document}
\title{Characterizing ultra-narrow momentum of atoms by standing-wave light-pulse sequences}
\author{ Shuyu Zhou }
\email{ syz@siom.ac.cn }
\affiliation{ Shanghai Institute of Optics and Fine Mechanics, Chinese Academy of Sciences, Shanghai 201800, China }
\affiliation{ Center of Materials Science and Optoelectronics Engineering, University of Chinese Academy of Sciences, Beijing 100049, China }
\author{ Chen Chen }
\affiliation{ Shanghai Institute of Optics and Fine Mechanics, Chinese Academy of Sciences, Shanghai 201800, China }
\affiliation{ Center of Materials Science and Optoelectronics Engineering, University of Chinese Academy of Sciences, Beijing 100049, China }
\author{ Bowen Xu }
\affiliation{ Shanghai Institute of Optics and Fine Mechanics, Chinese Academy of Sciences, Shanghai 201800, China }
\affiliation{ Center of Materials Science and Optoelectronics Engineering, University of Chinese Academy of Sciences, Beijing 100049, China }
\author{ Angang Liang }
\affiliation{ Shanghai Institute of Optics and Fine Mechanics, Chinese Academy of Sciences, Shanghai 201800, China }
\affiliation{ Center of Materials Science and Optoelectronics Engineering, University of Chinese Academy of Sciences, Beijing 100049, China }
\author{ Ying Wang }
\affiliation{ School of Science, Jiangsu University of Science and Technology, Zhenjiang 212003, China}
\author{Bin Wang }
\affiliation{ Shanghai Institute of Optics and Fine Mechanics, Chinese Academy of Sciences, Shanghai 201800, China }

\begin{abstract}
We propose a method to characterize the ultra-narrow momentum distribution of atomic gases by employing a standing-wave light-pulse sequences beam-splitter. The mechanism of beam splitting is analyzed in detail, and the influence of a finite-width momentum distribution on the population of each diffraction order is given. The temperature of ultracold atomic gases can be calibrated by measuring the ratio of population in different diffraction orders after double standing-wave light-pulses. We obtain analytical expressions for two typical cases, and demonstrate phase space evolution in the whole process by using the Wigner function. This method is valid for both classical atomic gas and Bose-Einstein condensates, and it is suited for temperature measurement on the space ultra-cold atomic physics platform, in which the ultra-narrow momentum distribution of atomic gas is on the order of $100 pK$ or even lower.
\end{abstract}

\maketitle

\section{I. Introduction}

Atomic samples with ultralow kinetic energy provide a favorable experimental condition for basic research and high-tech applications such as testing quantum mechanics at macroscopic scales \cite{001,002}, preparing the exotic phases of matter \cite{003,004,005,006}, exploring few-body physics \cite{007,008,009}, and developing high-precision atomic interferometer \cite{010,011}. In principle, ultracold atomic samples with kinetic energy of $100 pK$ or less can be obtained by adiabatic decompression \cite{012,013} or delta kick cooling \cite{014,015,016}. In the microgravity environment of the international space station, the kinetic energy equivalent temperature of Bose-Einstein condensates (BEC) was reduced to a minimum of about  $230 pK$ after adiabatic expansion \cite{017}. Recently, by combining an interaction-driven quadrupole-mode excitation of a BEC with a magnetic lens, the total internal kinetic energy of a BEC in three dimensions was lowered to $38 pK$ \cite{018}.
\\
\indent
Common approaches of measuring temperatures or momentum distributions of ultracold atoms include the time-of-flight (TOF) method \cite{017,018} and Bragg spectroscopy \cite{019,020}. The atomic knife-edge method has also been used for measuring narrow momentum distributions \cite{021}. Nevertheless, when we consider a realistic ultracold atomic experimental device, which is limited by the space station, we will find all above methods have their own weaknesses to measure such low temperature. We discuss this problem in detail in Section V.
\\
\indent
The finite momentum width of atoms will affect the contrast of the interference fringes \cite{022,023}. Based on the two-photon Raman transition method, temperature measurements have been demonstrated for a cold atom ensemble \cite{024} and a single atom \cite{025}. However, this scheme also requires precise control of the laser phase and intensity, and because Raman transitions involve different internal states, it is susceptible to the effects of magnetic field inhomogeneity and fluctuation.
\\
\indent
In this paper, we suggest using the method of standing-wave light-pulse sequences \cite{026,027} to measure the temperature of ultracold atoms or the kinetic energy equivalent temperature of BECs. The method we propose is well suited for ultracold atoms or BECs with kinetic energy below  $1 nK$, and can even be lower than $100 pK$. This approach is easy to realize since it is based on the common one-dimensional optical lattice and only needs appropriate timing control.
\\
\indent
This paper is organized as follows. In Sec. II, we discuss the mechanism of double-pulse beam splitting and establish the theoretical framework. Section III considers the effect of a finite width momentum distribution and delivers the analytical expressions of the population of diffraction orders verse the momentum width under two typical cases. The phase space evolution in the beam-splitting process is presented in Secs. IV., and the physical image of the diffractive order intensity change caused by the momentum broadening is discussed. In sec.V, we compare our proposed method with other methods for measuring temperature of ultracold atoms, discuss the effects due to interaction between atoms and vibrations in the space station, and illustrate the advantage of shorter wavelengths of optical lattice laser in experimental implementation. Finally, a brief summary is given in Sec. VI.

\section{II. Mechanism of double-pulse beam-splitting}
In this section we discuss the mechanism of double-pulse beam splitting. The potential of the one-dimensional optical lattice is
\begin{equation}
V\left( x \right) = 2\hbar \Omega _m \cos ^2 \left( {k_0 x} \right) = sE_r \cos ^2 \left( {k_0 x} \right).
\end{equation}
where $E_r {\rm{ = }}\hbar \omega _r$ is the recoil energy and $\omega _r  = \frac{{\hbar k_0^2 }}{{2m}}$ is the angular frequency corresponding to the photon recoil energy, $m$ is the mass of the atom and $k_0$ is the mode of the wave vector of the light field.
\\
\indent
The method of beam-splitting is described as two standing wave square pulses of amplitude $\Omega _m  = 2\sqrt 2 \omega _r$, both of pulse width $\tau _1  = \frac{{\left( {2n_1  + 1} \right)\pi }}{{4\sqrt 2 \omega _r }}$ and the interval time $\tau _2  = \frac{{\left( {2n_2  + 1} \right)\pi }}{{4\omega _r }}$, as shown in Fig.1(a) \cite{026}.
\\
\indent
\begin{figure}[h!]
\centering\includegraphics[width=7cm]{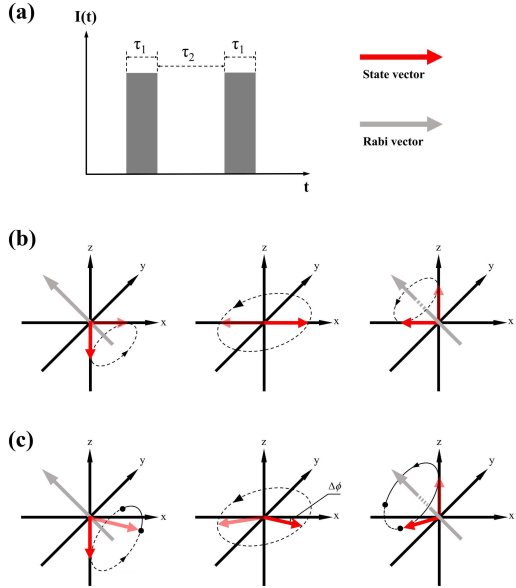}
\caption{Schematic diagram of double standing wave pulse beam splitting. (a) The sequence of the two-pulse. (b) The Bloch sphere interpretation of the matter-wave evolution. (a)(b) is similar to Fig. 1 in Ref. \cite{026}. (c) Evolution of the state vector in the Bloch sphere when the trap depth of the optical lattice is slightly higher than the optimal value.}
\end{figure}
\\
\indent
In Ref. \cite{026}, the motion of the atom was described by the coupled Raman-Nath equations (RNEs) and the RNEs was cut off to include only the $n = 0$ and $\pm 1$ diffraction orders. The $n =  \pm 1$ diffraction order forms a symmetric state $\left| {C_ +  } \right\rangle  = \frac{1}{{\sqrt 2 }}\left( {\left| {C_{ + 2} } \right\rangle  + \left| {C_{ - 2} } \right\rangle } \right)]$, which together with the zero-momentum state $\left| {C_0 } \right\rangle$ forms a two-state system. The evolution process of the double-pulse beam splitting can be expressed intuitively by the rotation of the state vector on a Bloch sphere, as shown in Fig.1(b). On the Bloch sphere, the state vector rotates around the Rabi vector $\mathord{\buildrel{\lower3pt\hbox{$\scriptscriptstyle\rightharpoonup$}} 
\over \Omega } {\rm{ = }}\left( {\sqrt 2 \Omega _m ,0,4\omega _r } \right)]$, which is caused by the standing wave light pulse. The state vector rotating from the $- Z$ axis to the $X$ axis or from the $-X$ axis to the $Z$ axis is equivalent to a Hadmard gate operation, while the rotation of the state vector from the $X$ axis to the $-X$ axis in the free evolutionary time is equivalent to the function of a $\pi$ phase gate. It is well known that the effect of sequential operation process of a Hadmard gate, a $\pi$ phase gate and a Hadmard gate is equivalent to the non-gate operation, or bit reversal. Thus, the atoms are initially in the zero-momentum state $\left| {C_0 } \right\rangle$, and after this sequence of operations will evolve to the symmetric state $\left| {C_ +  } \right\rangle  = \frac{1}{{\sqrt 2 }}\left( {\left| {C_{ + 2} } \right\rangle  + \left| {C_{ - 2} } \right\rangle } \right)]$.
\\
\indent
The above physical picture is clear, but there are two remaining questions. The first question is: can we regard the light pulse as doing phase imprinting on atoms? The reason we ask this question is that in this scheme, the individual pulses are only tens of microseconds long, and the movement of atoms during this time can probably be ignored. Usually, in this case, we could think of the light pulse as doing phase imprinting on atoms.
\\
\indent
The shortest time between two standing wave pulses is   
\begin{equation}
\tau _2  = \frac{\pi }{{4\omega _r }} = \frac{\pi }{4}\frac{{2m}}{{\hbar k_0^2 }} = \frac{m}{{2\hbar k_0 }}\frac{\pi }{{k_0 }} = \frac{d}{{v_{recoil} }},
\end{equation}
here $d{{ = \lambda } \mathord{\left/{\vphantom {{ = \lambda } 2}} \right.\kern-\nulldelimiterspace} 2}$ is the grating constant. Interestly, the Talbot length for the secondary Talbot image is $z_T  = \frac{{d^2 }}{\lambda }$, so the time it takes for light to travel this distance is
\begin{equation}
\tau _T  = \frac{{z_T }}{c} = \frac{1}{c}\frac{d}{\lambda }d = \frac{1}{{c\sin \theta _1 }}d = \frac{d}{{v_{recoil} }}.
\end{equation}
Therefore, the minimum propagation time in the double-pulse beam splitting is the same expression of as the time required for light to propagate from the grating to the Talbot length for the secondary Talbot image position \cite{028,029}. Can we consider the double-pulse beam splitting as doing phase imprinting twice and the traveling distance between the two pulses is the Talbot length for the secondary Talbot image? We will prove that this perception is incorrect.
\\
\indent
If the initial wavefront is
\begin{equation}
A\left( x \right) = \sum\limits_{n =  - \infty }^\infty  {C_n } e^{\frac{{i2\pi nx}}{d}},
\end{equation}
here $d$ is the grating constant. After traveling the distance z, the wave front is \cite{029}
\begin{equation}
E\left( x \right) \propto \sum\limits_{n =  - \infty }^\infty  {C_n } e^{ - i\frac{{\pi \lambda n^2 z}}{{d^2 }}} e^{i\frac{{2\pi nx}}{d}}.
\end{equation}
We assume that we start with a plane wave of unit amplitude, and add a phase-only modulation $e^{iz\cos \frac{{2\pi x}}{d}}$, then the output wavefront is the same as the amplitude transmittance
\begin{equation}
T\left( x \right) = \exp \left[ {iz\cos \frac{{2\pi x}}{d}} \right] = \sum\limits_{n =  - \infty }^\infty  {i^n } J_n \left( z \right)e^{i\frac{{2\pi nx}}{d}} {\rm{ = }}B{\rm{ + }}C.
\end{equation}
Here we denote the sum of the even terms as $B$ and the sum of the odd terms as $C$.
\\
\indent
According to Eq. (5), after the wave propagate a distance $d$, the wavefront is
\begin{equation}
E\left( x \right) \propto \sum\limits_{n =  - \infty }^\infty  {i^n } \left( { - 1} \right)^{n^2 } J_n \left( z \right)e^{i\frac{{2\pi nx}}{d}}.
\end{equation}
Compared with $\sum\limits_{n =  - \infty }^\infty  {i^n } J_n \left( z \right)e^{i\frac{{2\pi nx}}{d}}$, $\sum\limits_{n =  - \infty }^\infty  {i^n } \left( { - 1} \right)^{n^2 } J_n \left( z \right)e^{i\frac{{2\pi nx}}{d}}$ is characterized by the fact that  even terms do not change, and odd terms are multiplied by $-1$. So we have
\begin{equation}
E\left( x \right) \propto B - C.
\end{equation}
Then we do the phase - only modulation $e^{iz\cos \frac{{2\pi x}}{d}} $ again, the final wavefront is
\begin{equation}
E\left( x \right)T\left( x \right) \propto B^2  - C^2  = 1.
\end{equation}
So the outgoing wave is still a plane wave. This indicates that double-pulse beam splitting cannot be regarded as double phase imprinting of a thin grating.
\\
\indent
Then the second question rises: whether the standing wave pulse can be treated as a Hadmard gate under more strict conditions, rather than make an approximation including only the lowest-order Raman-Nath equations?
\\
\indent
To answer this question, we use the Bloch wave band method to deal with the effect of square pulse on atoms, which is also the approach to studying the Kapitza-Dirac Scattering. Bloch wave functions $\psi _{n,q} \left( x \right)$ are the products of a plane wave $e^{iqk_0 x}$ and a function $u_{n,q} \left( x \right)$ with the periodicity of the optical lattice
\begin{equation}
\psi _{n,q} \left( x \right) = e^{iqk_0 x} u_{n,q} \left( x \right),
\end{equation}
where
\begin{equation}
u_{n,q} \left( x \right) = \sum\limits_{l =  - \infty }^\infty  {c_{l,n,q} } e^{i2lk_0 x}.
\end{equation}
To deal with the evolution in optical lattices, the key is to find coefficients $c_{l,n,q}$ and eigenenergies $\varepsilon _{n,q}$ of $u_{n,q} \left( x \right)$. We can get them by numerically solving the matrix equation \cite{030}
\begin{equation}
\sum\limits_{l =  - \infty }^\infty  {H_{l,l'}  \cdot c_{l,n,q} }  = \varepsilon _{n,q} c_{l,n,q'},
\end{equation}
where
\begin{equation}
H_{l,l'}  = \left\{ {\begin{array}{*{20}c}
   {\left( {2l + q} \right)^2  + {s \mathord{\left/
 {\vphantom {s {2\begin{array}{*{20}c}
   {} & {if} & {} & {l = l'}  \\
\end{array}}}} \right.
 \kern-\nulldelimiterspace} {2\begin{array}{*{20}c}
   {} & {if} & {} & {l = l'}  \\
\end{array}}}}  \\
   { - {s \mathord{\left/
 {\vphantom {s {4\begin{array}{*{20}c}
   {} & {if} & {} & {\left| {l - l'} \right| = 1}  \\
\end{array}}}} \right.
 \kern-\nulldelimiterspace} {4\begin{array}{*{20}c}
   {} & {if} & {} & {\left| {l - l'} \right| = 1}  \\
\end{array}}}}  \\
   {0\begin{array}{*{20}c}
   {} & {else} & {} & {}  \\
\end{array}}  \\
\end{array}} \right.{\rm{.}}
\end{equation}
When atoms are suddenly loaded into an optical lattice, they are projected onto superpositions of different Bloch bands. For the case where $q = 0$ and the potential-well depth of the optical lattice is relatively shallow, for example when $s < 10$, only bands $0$ and $2$ have significant populations, and the components with momentum $0$ and  $\pm 2\hbar k_0$ account for the majority of these bands \cite{031}. In our actual calculation, we truncated the equation from $l =  - 20$ to $l = 20$, so the calculation accuracy is high enough. Our calculation strategy is to increase the well depth from a small value, until the minimum of the population of zero momentum part at the first oscillation period approaches $0.5$. The population of $+ 2\hbar k_0$ and $- 2\hbar k_0$ parts at this time will be very close to $0.25$, while the population of the other momentum components can be ignored. Here we choose the population of zero momentum part equaling $0.5$ at the minimum point of the first oscilltion period, which suggests that such a pulse does indeed rotate the state vector from $-Z$ to $X$ axes and a subsequent standing wave pulse of the same length can return the atom to its initial zero-momentum state. Therefore, this standing wave square optical pulse can realize a Hadamard gate operation well.
\\
\indent
For example, if the wavelength of the lattice beam is $780 nm$ and the initial momentum is zero, when $s = 5.5$ and the pulse width $\tau _1  = 24.8\mu s$, the population of the zero momentum part is $0.4984$, and the population of $+ 2\hbar k_0$ and $- 2\hbar k_0$ momentum part are both $0.2490$, which is close to a good Hadamard gate operation.
\\
\indent
If the initial momentum is not strictly zero, but a small momentum $q\hbar k_0$ ($\left| q \right| <  < 1$), the standing wave square optical pulse can still perform an Hadamard gate operation approximately. For example, if the wavelength of the lattice beam is $532 nm$ and the initial momentum is $0.03\hbar k_0$, when $s = 5.5$ and the pulse width $\tau _1  = 11.7\mu s$, the populations of order zero and order $\pm 1$ diffraction are $0.4996$, $0.2391$ and $0.2576$, respectively. The phase difference between order zero and order $\pm 1$ diffraction varies by only $0.05$ radian compared to the case where the initial momentum is zero. Besides, a subsequent standing wave pulse of the same length still can return the atom almost completely to its initial zero-momentum state, as showed in Fig. 2.
\\
\indent
\begin{figure}[h!]
\centering\includegraphics[width=7cm]{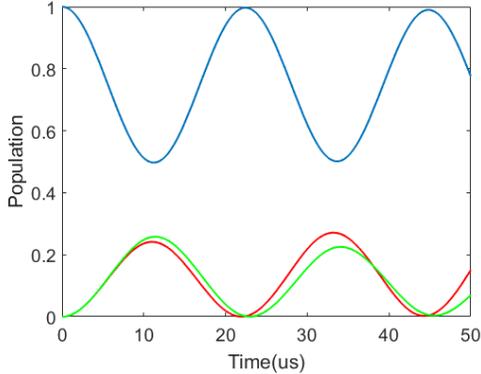}
\caption{Population of momentum components as a function of lattice pulse time. The blue, red, and green lines correspond to populations with momentum $0\hbar k_0$, $2\hbar k_0$, and $- 2\hbar k_0$ parts, respectively.}
\end{figure}
\\
\indent
We would like to discuss the slight difference between the optimization in Ref. \cite{026,027} and our results. In Ref. \cite{026}, the amplitude of the double square pulse is fixed at $\Omega _m  = 2\sqrt 2 \omega _r$, which corresponds to $s = 5.66$ and is slightly above our set point. After optimization, the pulse width $\tau _1$ is shorter than $\frac{\pi }{{4\sqrt 2 \omega _r }}$ and the interval time $\tau _2$ is longer than $\frac{\pi }{{4\omega _r }}$. Here we will give an intuitive physical images to understand these results. The state vector rotates around the Rabi vector $\mathord{\buildrel{\lower3pt\hbox{$\scriptscriptstyle\rightharpoonup$}} 
\over \Omega } {\rm{ = }}\left( {\sqrt 2 \Omega _m ,0,4\omega _r } \right)$ is an approximate physical model by truncating the Raman-Nath equations to the lowest-order 3 equations.According to this model, if we set $\Omega _m  = 2\sqrt 2 \omega _r$ and the pulse length $\tau _1  = \frac{\pi }{{4\sqrt 2 \omega _r }}$, the state vector starting from the $-Z$ axis should stop at the $X$ axis, which is the top of the rotating track. Nevertheless, we calculated the population evolution of the part with zero momentum by using the Bloch wave band method with the same parameter. The result showed the population of zero momentum part is less than $0.5$ at the minimum point of the first oscillation period, which means the top of the rotating track is higher than the $X-Y$ plane, as shown in Fig. 1(c). If we want to stop the state vector on the $X-Y$ plane, the pulse length $\tau _1$ should be relatively shorter, which also brings in an additional phase $\Delta \phi$. In order for the next light pulse to rotate the state vector to the $Z$ axis, the phase change corresponding to the free evolution is not $\pi$ but $\pi {\rm{ + }}2\Delta \phi$, so it takes a longer time.
\\
\indent
Then we deduced the evolution matrix corresponding to free propagation and presented the whole theoretical framework. We can choose the non-coupled basis vectors $\left\{ {\left| {C_0 } \right\rangle ,\left| {C_{ + 2} } \right\rangle ,\left| {C_{ - 2} } \right\rangle } \right\}$, where $\left| {C_0 } \right\rangle$, $\left| {C_{ + 2} } \right\rangle$ and $\left| {C_{ - 2} } \right\rangle$ represent states with momentum $q\hbar k_0$, $\left( {2 + q} \right)\hbar k_0$ and $\left( { - 2 + q} \right)\hbar k_0$, respectively. In the momentum representation, their wave functions are $\delta \left( {p' - \hbar k} \right)$, $\delta \left( {p' - \left( {\hbar k - 2\hbar k_0 } \right)} \right)$ and $\delta \left( {p' - \left( {\hbar k + 2\hbar k_0 } \right)} \right)$. We can also choose the coupled basis vectors $\left\{ {\left| {C_0 } \right\rangle ,\left| {C_ +  } \right\rangle ,\left| {C_ -  } \right\rangle } \right\}$, where $\left| {C_ +  } \right\rangle  = \frac{1}{{\sqrt 2 }}\left( {\left| {C_{ + 2} } \right\rangle  + \left| {C_{ - 2} } \right\rangle } \right)$ and $\left| {C_ -  } \right\rangle  = \frac{1}{{\sqrt 2 }}\left( {\left| {C_{ + 2} } \right\rangle  - \left| {C_{ - 2} } \right\rangle } \right)$. However, in order to make operations similar to the previous Hadmard gate - phase gate - Hadmard gate, we prefer to do it on a coupled basis vectors.
\\
\indent
After propagation, we take the phase change of state $\left| {C_0 } \right\rangle$ as the reference zero, then the phase change of $\left| {C_{ + 2} } \right\rangle$ is $\varphi _{ + 2}$, and the phase change of $\left| {C_{ - 2} } \right\rangle$ is $\varphi _{ - 2}$. The state $\left| {C_{ + 2} } \right\rangle$ has higher energy than $\left| {C_0 } \right\rangle$ by
\begin{equation}
\Delta E_{ + 2}  = \frac{{\hbar ^2 \left( {2 + q} \right)^2 k_0^2 }}{{2m}} - \frac{{\hbar ^2 q^2 k_0^2 }}{{2m}} = \frac{{\left( {2 + 2q} \right)\hbar ^2 k_0^2 }}{m},
\end{equation}
so the phase difference between $\left| {C_{ + 2} } \right\rangle$ and $\left| {C_0 } \right\rangle$ is
\begin{equation}
\varphi _{ + 2}  = \frac{{\Delta E_{ + 2} \tau _2 }}{\hbar } = \frac{{\left( {2 + 2q} \right)\hbar k_0^2 \tau _2 }}{m},
\end{equation}
where $\tau _2$ is the time of free evolution between two pulses. We can set $\varphi _{ + 2}  = \theta  + \Delta \theta$, where $\theta  = \frac{{2\hbar k_0^2 \tau _2 }}{m}$ is the phase difference corresponding $q = 0$, and $\Delta \theta  = \frac{{2q\hbar k_0^2 \tau _2 }}{m}$ is the contribution from $q$. The phase difference between $\left| {C_{ - 2} } \right\rangle$ and $\left| {C_0 } \right\rangle$ is
\begin{equation}
\varphi _{ - 2}  = \frac{{\Delta E_{ - 2} \tau _2 }}{\hbar } = \frac{{\left( {2 - 2q} \right)\hbar k_0^2 \tau _2 }}{m}.
\end{equation}
Similarly, $\varphi _{ - 2}$ can be expressed as $\varphi _{ - 2}  = \theta  - \Delta \theta$. We know that the momentum eigenstate satisfies
\begin{equation}
\Phi \left( {k,t} \right) = \Phi \left( {k,0} \right)\exp \left[ {{{ - iEt} \mathord{\left/
 {\vphantom {{ - iEt} \hbar }} \right.
 \kern-\nulldelimiterspace} \hbar }} \right] = \Phi \left( {k,0} \right)\exp \left[ { - i\varphi } \right].
\end{equation}
Then, under the representation with $\left\{ {\left| {C_0 } \right\rangle ,\left| {C_{ + 2} } \right\rangle ,\left| {C_{ - 2} } \right\rangle } \right\}$ as base kets, the evolution matrix of free propagation is
\begin{equation}
F = \left[ {\begin{array}{*{20}c}
   1 & 0 & 0  \\
   0 & {e^{ - i\varphi _{ + 2} } } & 0  \\
   0 & 0 & {e^{ - i\varphi _{ - 2} } }  \\
\end{array}} \right] = \left[ {\begin{array}{*{20}c}
   1 & 0 & 0  \\
   0 & {e^{ - i\left( {\theta  + \Delta \theta } \right)} } & 0  \\
   0 & 0 & {e^{ - i\left( {\theta  - \Delta \theta } \right)} }  \\
\end{array}} \right].
\end{equation}
If we chose base kets as $\left\{ {\left| {C_0 } \right\rangle ,\left| {C_ +  } \right\rangle ,\left| {C_ -  } \right\rangle } \right\}$, the evolution matrix is $P = S^ +  FS$, where representation transformation matrix
\begin{equation}
S = S^ +   = S^{ - 1}  = \left[ {\begin{array}{*{20}c}
   1 & 0 & 0  \\
   0 & {\frac{1}{{\sqrt 2 }}} & {\frac{1}{{\sqrt 2 }}}  \\
   0 & {\frac{1}{{\sqrt 2 }}} & { - \frac{1}{{\sqrt 2 }}}  \\
\end{array}} \right].
\end{equation}
Therefore, the evolution matrix of free propagation under the representation with base kets $\left\{ {\left| {C_0 } \right\rangle ,\left| {C_{ + 2} } \right\rangle ,\left| {C_{ - 2} } \right\rangle } \right\}$ is
\begin{equation}
P = \left[ {\begin{array}{*{20}c}
   1 & 0 & 0  \\
   0 & {e^{ - i\theta } \cos \Delta \theta } & { - ie^{ - i\theta } \sin \Delta \theta }  \\
   0 & { - ie^{ - i\theta } \sin \Delta \theta } & {e^{ - i\theta } \cos \Delta \theta }  \\
\end{array}} \right].
\end{equation}
\\
\indent
Next, we will examine the matrix corresponding to the standing wave square pulse. By selecting the appropriate parameters, we make the evolution matrix corresponding to the pulse under the representation of $\left\{ {\left| {C_0 } \right\rangle ,\left| {C_ +  } \right\rangle ,\left| {C_ -  } \right\rangle } \right\}$  be
\begin{equation}
H = \left[ {\begin{array}{*{20}c}
   {\frac{1}{{\sqrt 2 }}} & {\frac{1}{{\sqrt 2 }}} & 0  \\
   {\frac{1}{{\sqrt 2 }}} & { - \frac{1}{{\sqrt 2 }}} & 0  \\
   0 & 0 & 1  \\
\end{array}} \right].
\end{equation}
This matrix implies that the symmetric basis vectors $\left| {C_0 } \right\rangle$ and $\left| {C_ +  } \right\rangle$ are decoupled from the asymmetric basis vector $\left| {C_ -  } \right\rangle$, and a Hadamard gate operation is performed for $\left| {C_0 } \right\rangle$ and $\left| {C_ +  } \right\rangle$.
\\
\indent
Thus, the unitary transformation corresponding to the complete impulse - free evolution - impulse ($HPH$) is
\begin{equation}
HPH = \left[ {\begin{array}{*{20}c}
   {\frac{{1 + e^{ - i\theta } \cos \theta }}{2}} & {\frac{{1 - e^{ - i\theta } \cos \theta }}{2}} & { - \frac{{ie^{ - i\theta } \sin \theta }}{{\sqrt 2 }}}  \\
   {\frac{{1 - e^{ - i\theta } \cos \theta }}{2}} & {\frac{{1 + e^{ - i\theta } \cos \theta }}{2}} & {\frac{{ie^{ - i\theta } \sin \theta }}{{\sqrt 2 }}}  \\
   { - \frac{{ie^{ - i\theta } \sin \theta }}{{\sqrt 2 }}} & {\frac{{ie^{ - i\theta } \sin \theta }}{{\sqrt 2 }}} & {e^{ - i\theta } \cos \theta }  \\
\end{array}} \right].
\end{equation}
If the momentum of the initial state is $q\hbar k_0$, then
\begin{equation}
HPH\left[ {\begin{array}{*{20}c}
   1  \\
   0  \\
   0  \\
\end{array}} \right] = \left[ {\begin{array}{*{20}c}
   {\frac{{1 + e^{ - i\theta } \cos \Delta \theta }}{2}}  \\
   {\frac{{1 - e^{ - i\theta } \cos \Delta \theta }}{2}}  \\
   { - \frac{{ie^{ - i\theta } \sin \Delta \theta }}{{\sqrt 2 }}}  \\
\end{array}} \right].
\end{equation}
For the case of $\theta  = \left( {2N + 1} \right)\pi$ and $q = 0$, the final output is $\left| {C_ +  } \right\rangle$, this is exactly the case of Ref. \cite{026}.

\section{III. The effect of finite momentum width}

Next, we will consider the effect of a finite width momentum distribution. Eq. (23) already gives the final state under the representation of $\left\{ {\left| {C_0 } \right\rangle ,\left| {C_ +  } \right\rangle ,\left| {C_ -  } \right\rangle } \right\}$ when the initial state momentum is $q\hbar k_0$. Using the representation transformation formula, we can obtain the final state output under representation of $\left\{ {\left| {C_0 } \right\rangle ,\left| {C_{ + 2} } \right\rangle ,\left| {C_{ - 2} } \right\rangle } \right\}$ is
\begin{equation}
\left[ {\begin{array}{*{20}c}
   1 & 0 & 0  \\
   0 & {\frac{1}{{\sqrt 2 }}} & {\frac{1}{{\sqrt 2 }}}  \\
   0 & {\frac{1}{{\sqrt 2 }}} & { - \frac{1}{{\sqrt 2 }}}  \\
\end{array}} \right]\left[ {\begin{array}{*{20}c}
   {\frac{{1 + e^{ - i\theta } \cos \Delta \theta }}{2}}  \\
   {\frac{{1 - e^{ - i\theta } \cos \Delta \theta }}{2}}  \\
   { - \frac{{ie^{ - i\theta } \sin \Delta \theta }}{{\sqrt 2 }}}  \\
\end{array}} \right] = \left[ {\begin{array}{*{20}c}
   {\frac{{1 + e^{ - i\theta } \cos \Delta \theta }}{2}}  \\
   {\frac{{\left( {1 - e^{ - i\theta } \cos \Delta \theta } \right) - ie^{ - i\theta } \sqrt 2 \sin \Delta \theta }}{{2\sqrt 2 }}}  \\
   {\frac{{\left( {1 - e^{ - i\theta } \cos \Delta \theta } \right) + ie^{ - i\theta } \sqrt 2 \sin \theta }}{{2\sqrt 2 }}}  \\
\end{array}} \right].
\end{equation}
The wave function under the momentum representation can be expanded as
\begin{equation}
\Phi \left( p \right) = \int {\Phi \left( {p'} \right)} \delta \left( {p' - p} \right)dp',
\end{equation}
where $\delta \left( {p' - p} \right) = \delta \left( {p - p'} \right)$ is the momentum eigenstate with the eigenvalue of $p$, and the expansion coefficient is $\Phi \left( {p'} \right)$. If the initial wave function is
\begin{equation}
\Phi _{ini} \left( {p = \hbar k} \right) = \int {\Phi \left( {p'} \right)} \delta \left( {p' - \hbar k} \right)dp',
\end{equation}
after $HPH$ evolution, $\delta \left( {p' - q\hbar k_0 } \right)$ evolves into
\begin{equation}
\begin{array}{l}
 \frac{{1 + e^{ - i\theta } \cos \Delta \theta }}{2}\delta \left( {p' - \hbar k} \right) + \frac{{\left( {1 - e^{ - i\theta } \cos \Delta \theta } \right) - ie^{ - i\theta } \sqrt 2 \sin \Delta \theta }}{{2\sqrt 2 }}\delta \left( {p' - \left( {\hbar k - 2\hbar k_0 } \right)} \right) \\ 
  + \frac{{\left( {1 - e^{ - i\theta } \cos \Delta \theta } \right) + ie^{ - i\theta } \sqrt 2 \sin \Delta \theta }}{{2\sqrt 2 }}\delta \left( {p' - \left( {\hbar k + 2\hbar k_0 } \right)} \right) \\ 
 \end{array}.
\end{equation}
So the final wave function is
\begin{equation}
\Phi _{fin} \left( {p = \hbar k} \right) = \int {\Phi \left( {p'} \right)} \left[ \begin{array}{l}
 \frac{{1 + e^{ - i\theta } \cos \Delta \theta }}{2}\delta \left( {p' - \hbar k} \right) + \frac{{\left( {1 - e^{ - i\theta } \cos \Delta \theta } \right) - ie^{ - i\theta } \sqrt 2 \sin \Delta \theta }}{{2\sqrt 2 }}\delta \left( {p' - \left( {\hbar k - 2\hbar k_0 } \right)} \right) \\ 
  + \frac{{\left( {1 - e^{ - i\theta } \cos \Delta \theta } \right) + ie^{ - i\theta } \sqrt 2 \sin \Delta \theta }}{{2\sqrt 2 }}\delta \left( {p' - \left( {\hbar k + 2\hbar k_0 } \right)} \right) \\ 
 \end{array} \right]dp'.
\end{equation}
If $\left| q \right| <  < 1$, the central zero-momentum term and the $\pm 2\hbar k_0$ terms are separated after a time-of-flight expansion in free space. The information about the momentum distribution can be obtained by statistical population of the terms. 
\\
\indent
As an example, we take the wave function of the initial state as a least uncertain Gaussian wave packet
\begin{equation}
\Phi _{ini} \left( k \right) = \frac{1}{{\alpha ^{{1 \mathord{\left/
 {\vphantom {1 2}} \right.
 \kern-\nulldelimiterspace} 2}} \pi ^{{1 \mathord{\left/
 {\vphantom {1 4}} \right.
 \kern-\nulldelimiterspace} 4}} }}e^{ - \frac{{k^2 }}{{2\alpha ^2 }}} .
\end{equation}
If we set $k = qk_0$ and $\alpha  = q_m k_0$, the width of the initial wave packet in momentum space in units $k_0$ is equal to $q_m$, the normalized initial state wave function in terms of $q$ is
\begin{equation}
\Phi _{ini} \left( q \right) = \frac{1}{{q_m^{{1 \mathord{\left/
 {\vphantom {1 2}} \right.
 \kern-\nulldelimiterspace} 2}} \pi ^{{1 \mathord{\left/
 {\vphantom {1 4}} \right.
 \kern-\nulldelimiterspace} 4}} }}e^{ - \frac{{q^2 }}{{2q_m^2 }}}.
\end{equation}
The final wave function is
\begin{equation}
\begin{array}{l}
 \Phi _{final} \left( k \right) = \frac{1}{{\alpha ^{{1 \mathord{\left/
 {\vphantom {1 2}} \right.
 \kern-\nulldelimiterspace} 2}} \pi ^{{1 \mathord{\left/
 {\vphantom {1 4}} \right.
 \kern-\nulldelimiterspace} 4}} }}e^{ - \frac{{k^2 }}{{2\alpha ^2 }}}  \cdot \frac{{1 + e^{ - i\theta } \cos \Delta \theta }}{2} + \frac{1}{{\alpha ^{{1 \mathord{\left/
 {\vphantom {1 2}} \right.
 \kern-\nulldelimiterspace} 2}} \pi ^{{1 \mathord{\left/
 {\vphantom {1 4}} \right.
 \kern-\nulldelimiterspace} 4}} }}e^{ - \frac{{\left( {k - 2k_0 } \right)^2 }}{{2\alpha ^2 }}}  \cdot \frac{{\left( {1 - e^{ - i\theta } \cos \Delta \theta } \right) - ie^{ - i\theta } \sqrt 2 \sin \Delta \theta }}{{2\sqrt 2 }} \\ 
  + \frac{1}{{\alpha ^{{1 \mathord{\left/
 {\vphantom {1 2}} \right.
 \kern-\nulldelimiterspace} 2}} \pi ^{{1 \mathord{\left/
 {\vphantom {1 4}} \right.
 \kern-\nulldelimiterspace} 4}} }}e^{ - \frac{{\left( {k + 2k_0 } \right)^2 }}{{2\alpha ^2 }}}  \cdot \frac{{\left( {1 - e^{ - i\theta } \cos \Delta \theta } \right) + ie^{ - i\theta } \sqrt 2 \sin \Delta \theta }}{{2\sqrt 2 }} \\ 
 \end{array}.
\end{equation}
After sufficient free flight, the three parts of central momentum $0\hbar k_0$, $2\hbar k_0$, and $- 2\hbar k_0$ separate in space, with no interference terms. Therefore, the optical depth of these three parts can be experimentally integrated respectively in space, so as to obtain the proportion of these three parts. Wherein, the modulus square of the central zero-momentum term is
\begin{equation}
\frac{1}{{\alpha \sqrt \pi  }}e^{ - \frac{{k^2 }}{{\alpha ^2 }}}  \cdot \left( {\frac{{1 + e^{ - i\theta } \cos \Delta \theta }}{2}} \right)\left( {\frac{{1 + e^{i\theta } \cos \Delta \theta }}{2}} \right) = \frac{1}{{\alpha \sqrt \pi  }}e^{ - \frac{{k^2 }}{{\alpha ^2 }}} \left( {\frac{3}{8} + \frac{{\cos \theta }}{2}\cos \Delta \theta  + \frac{{\cos 2\Delta \theta }}{8}} \right).
\end{equation}
Then we integrate with respect to $k = qk_0$, and notice that $\Delta \theta  = \frac{{2q\hbar k_0^2 \tau _2 }}{m}$ contains the integral variable. For simplicity, we define a dimensionless time $t_2  = \frac{{2\hbar k_0^2 \tau _2 }}{{m\pi }}$, then $\Delta \theta  = t_2 \pi q$ and $\theta  = t_2 \pi$. The integral is
\begin{equation}
\int_{ - \infty }^{ + \infty } {\frac{1}{{\alpha \sqrt \pi  }}e^{ - \frac{{k^2 }}{{\alpha ^2 }}} \left( {\frac{3}{8} + \frac{{\cos \theta }}{2}\cos \Delta \theta  + \frac{{\cos 2\Delta \theta }}{8}} \right)} dk = \frac{3}{8} + \frac{1}{2}\cos \left( {t_2 \pi } \right)e^{ - \frac{{\pi ^2 q_m^2 t_2^2 }}{4}}  + \frac{1}{8}e^{ - \pi ^2 q_m^2 t_2^2 }.
\end{equation}
Similarly, we get that the population of the momentum part of $2\hbar k_0$ and $-2\hbar k_0$ both are $\frac{5}{{16}} - \frac{1}{4}\cos \left( {t_2 \pi } \right)e^{ - \frac{{\pi ^2 q_m^2 t_2^2 }}{4}}  - \frac{1}{{16}}e^{ - \pi ^2 q_m^2 t_2^2 }$. When $t_2  = 2N + 1$, where $N$ is an integer greater than or equal to zero, the ratio of central zero-momentum to $2\hbar k_0$ or $- 2\hbar k_0$ part is
\begin{equation}
Ratio = {{\left( {\frac{3}{8} - \frac{1}{2}e^{ - \frac{{\pi ^2 q_m^2 t_2^2 }}{4}}  + \frac{1}{8}e^{ - \pi ^2 q_m^2 t_2^2 } } \right)} \mathord{\left/
 {\vphantom {{\left( {\frac{3}{8} - \frac{1}{2}e^{ - \frac{{\pi ^2 q_m^2 t_2^2 }}{4}}  + \frac{1}{8}e^{ - \pi ^2 q_m^2 t_2^2 } } \right)} {\left( {\frac{5}{{16}} + \frac{1}{4}e^{ - \frac{{\pi ^2 q_m^2 t_2^2 }}{4}}  - \frac{1}{{16}}e^{ - \pi ^2 q_m^2 t_2^2 } } \right)}}} \right.
 \kern-\nulldelimiterspace} {\left( {\frac{5}{{16}} + \frac{1}{4}e^{ - \frac{{\pi ^2 q_m^2 t_2^2 }}{4}}  - \frac{1}{{16}}e^{ - \pi ^2 q_m^2 t_2^2 } } \right)}}.
\end{equation}
\indent
Although we have discussed the Gaussian wave packet of a single atom above, this method is still suitable for measuring the temperature of the ultracold atomic ensemble. Each atom in the ultracold atomic ensemble is treated as a Gaussian wave packet \cite{025,032}, and the width of the wave packet in the momentum space is determined by temperature $q_m^2  = \frac{{2k_B Tm}}{{k_0^2 \hbar ^2 }}$. The different initial positions of each atom will introduce a different initial phase on the momentum wave packet, but this initial phase will not affect the population of each diffraction order after $HPH$ evolution. As long as the flight time after $HPH$ evolution is sufficient, the diffraction order of the ultracold atomic cloud can be experimentally distinguished.
\\
\indent
Fig. 3(a) shows the evolution of the population of zero momentum term with the dimensionless time $t_2$ when the temperature is $300 pK$. Fig. 3(b) is the curve of the population ratio verse temperature when $t_2  = 31$.  In the above calculation, we set the wavelength of the optical lattice laser as $532 nm$, so the actual interval time $\tau _2$ is about $0.477 ms$. Therefore, even if the whole beamsplitting process of the dual standing wave pulse is less than $1 ms$, it is capable of detecting temperatures as low as $100 pK$ in principle.
\\
\indent
\begin{figure}[h!]
\centering\includegraphics[width=7cm]{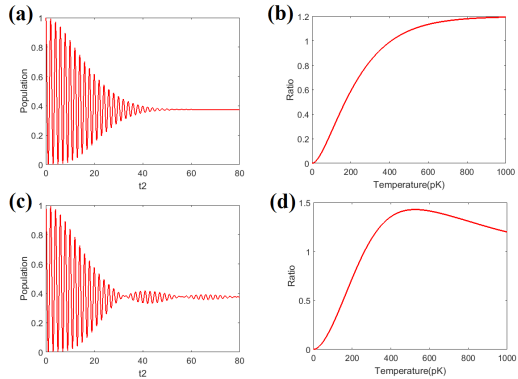}
\caption{The effect of finite momentum width. (a) Temporal evolution of the population of zero-momentum term with the dimensionless time $t_2$ when the temperature is $300 pK$. (b) The population ratio of $0\hbar k_0$ to $2\hbar k_0$ ($-2\hbar k_0$) verse temperature of the Gaussian wave packet when the dimensionless time $t_2  = 31$. (c) Temporal evolution of the population of zero-momentum term with the dimensionless time $t_2$ when the kinetic energy equivalent temperature of the BEC is $300 pK$. (d) The population ratio of $0\hbar k_0$ to $2\hbar k_0$ ($-2\hbar k_0$) verse the kinetic energy equivalent temperature of the BEC when the dimensionless time  $t_2  = 31$.}
\end{figure}
\\
\indent
Then we take another example of a BEC released from a trap. Under the Thomas-Fermi approximation, the density distribution of a BEC in the trap is \cite{033}
\begin{equation}
n\left( {\mathord{\buildrel{\lower3pt\hbox{$\scriptscriptstyle\rightharpoonup$}} 
\over r} } \right) = \frac{{\mu  - V\left( {\mathord{\buildrel{\lower3pt\hbox{$\scriptscriptstyle\rightharpoonup$}} 
\over r} } \right)}}{{U_0 }},
\end{equation}
Where $\mu$ is the chemical potential and $U_0$ is the effective interaction between two atoms. In an anisotropic three-dimensional harmonic-oscillator potential $V\left( {x,y,z} \right) = \frac{m}{2}\left( {\omega _x^2 x^2  + \omega _y^2 y^2  + \omega _z^2 z^2 } \right)$, the density distribution of the BEC is
\begin{equation}
n\left( {x,y,z} \right) = n\left( {0,0,0} \right)\max \left[ {1 - \frac{{x^2 }}{{R_x^2 }} - \frac{{y^2 }}{{R_y^2 }} - \frac{{z^2 }}{{R_z^2 }},0} \right],
\end{equation}
where $R_i  = \sqrt {\frac{{2\mu }}{{m\omega _i^2 }}} $ $\left( {i = x,y,z} \right)$ is the half-lengths of the trapped condensate. After released from the trap, the BEC evolves as a rescaling of its parabolic shape and the local velocity of the expanding cloud is \cite{034,035}
\begin{equation}
v_i \left( {\mathord{\buildrel{\lower3pt\hbox{$\scriptscriptstyle\rightharpoonup$}} 
\over r} ,t} \right) = r_i \frac{{\dot \lambda _i \left( t \right)}}{{\lambda _i \left( t \right)}},
\end{equation}
where $\lambda _i  = {{R_i \left( t \right)} \mathord{\left/
 {\vphantom {{R_i \left( t \right)} {R_i }}} \right.
 \kern-\nulldelimiterspace} {R_i }}\left( 0 \right)$ are the scaling factors. Therefore, if we selectively detect the region of $y \approx 0$ and $z \approx 0$ along the $X-$ axis, which can be achieved experimentally with a thin optical pumping beam, then we will find that the momentum distribution is also parabolic shape. So the initial wave function of the BEC in this region can be approximately expressed as
\begin{equation}
\Phi _{ini} \left( {p_x } \right) = Ae^{i\varphi \left( {p_x } \right)} \sqrt {1 - \frac{{p_x ^2 }}{{p_m^2 }}},
\end{equation}
here $p_x  = q\hbar k_0$, $p_m  = q_m \hbar k_0$ and $- p_m  \le p_x  \le p_m$. The wave function after $HPH$ evolution is
\begin{equation}
\begin{array}{l}
 \Phi _{final} \left( {p_x  = \hbar k} \right) = Ae^{i\varphi \left( {\hbar k} \right)} \sqrt {1 - \frac{{\left( {\hbar k} \right)^2 }}{{p_m^2 }}}  \cdot \frac{{1 + e^{ - i\theta } \cos \Delta \theta }}{2} \\ 
  + Ae^{i\varphi \left( {\hbar k - 2\hbar k_0 } \right)} \sqrt {1 - \frac{{\left( {\hbar k - 2\hbar k_0 } \right)^2 }}{{p_m^2 }}}  \cdot \frac{{\left( {1 - e^{ - i\theta } \cos \Delta \theta } \right) - ie^{ - i\theta } \sqrt 2 \sin \Delta \theta }}{{2\sqrt 2 }} \\ 
  + Ae^{i\varphi \left( {\hbar k + 2\hbar k_0 } \right)} \sqrt {1 - \frac{{\left( {\hbar k + 2\hbar k_0 } \right)^2 }}{{p_m^2 }}}  \cdot \frac{{\left( {1 - e^{ - i\theta } \cos \Delta \theta } \right) + ie^{ - i\theta } \sqrt 2 \sin \Delta \theta }}{{2\sqrt 2 }} \\ 
 \end{array},
\end{equation}
where the wave function of the zero-momentum term is $Ae^{i\varphi \left( {q\hbar k_0 } \right)} \sqrt {1 - \frac{{q^2 }}{{q_m^2 }}}  \cdot \frac{{1 + e^{ - i\theta } \cos \Delta \theta }}{2}$ and its population is $\int\limits_{ - q_m }^{q_m } B \left( {1 - \frac{{q^2 }}{{q_m^2 }}} \right) \cdot \frac{{1 + e^{ - i\theta } \cos \Delta \theta }}{2} \cdot \frac{{1 + e^{i\theta } \cos \Delta \theta }}{2}dq$.
Since the population of all atoms is unit, the normalization coefficient is $B = \frac{3}{{4q_m }}$, so the population of the zero momentum term is
\begin{equation}
\begin{array}{l}
 \int\limits_{ - q_m }^{q_m } {\frac{3}{{4q_m }}} \left( {1 - \frac{{q^2 }}{{q_m^2 }}} \right) \cdot \frac{{1 + e^{ - i\theta } \cos \Delta \theta }}{2} \cdot \frac{{1 + e^{i\theta } \cos \Delta \theta }}{2}dq \\ 
  = \frac{3}{8} + \frac{3}{{2\pi ^3 t_2^3 q_m^3 }}\cos \left( {\pi t_2 } \right) \cdot \sin \left( {\pi t_2 q_m } \right) - \frac{3}{{2\pi ^2 t_2^2 q_m^2 }}\cos \left( {\pi t_2 } \right) \cdot \cos \left( {\pi t_2 q_m } \right) \\ 
  + \frac{3}{{64\pi ^3 t_2^3 q_m^3 }}\sin \left( {2\pi t_2 q_m } \right) - \frac{3}{{32\pi ^2 t_2^2 q_m^2 }}\cos \left( {2\pi t_2 q_m } \right) \\ 
 \end{array}.
\end{equation}
Here $q_m^2  = \frac{{5k_B Tm}}{{k_0^2 \hbar ^2 }}$ is determined by the kinetic energy equivalent temperature $T$.
\\
\indent
We still set the wavelength of the optical lattice laser as $532 nm$. Fig.3(c) shows the evolution of the population of zero momentum term with dimensionless time $t_2$ when the equivalent temperature of the BEC is $300 pK$. And similarly, we got the curve of the population ratio verse equivalent temperature when $t_2  = 31$, as shown in Fig.3 (d). Comparing this curve with that of Gaussian wave packet (Fig.3(b)), we found there is an obvious deviation when the population ratio is bigger than $0.5$. Therefore, in the absence of prior information on the momentum distribution, it is best to limit the experimental test results to a population ratio less than $0.5$.

\section{IV. Phase space evolution of double-pulse beam-splitting process}

A quasi-probability distribution in phase space can be obtained by using the Wigner function \cite{036,037,038}, which provides a convenient physical interpretation for the double pulse beam splitting process. The Wigner function constructed based on the momentum representation wave function is \cite{039}
\begin{equation}
W\left( {\mathord{\buildrel{\lower3pt\hbox{$\scriptscriptstyle\rightharpoonup$}} 
\over x} ,\mathord{\buildrel{\lower3pt\hbox{$\scriptscriptstyle\rightharpoonup$}} 
\over k} } \right) = \int {e^{i\mathord{\buildrel{\lower3pt\hbox{$\scriptscriptstyle\rightharpoonup$}} 
\over p} \mathord{\buildrel{\lower3pt\hbox{$\scriptscriptstyle\rightharpoonup$}} 
\over x} } \phi \left( { - \mathord{\buildrel{\lower3pt\hbox{$\scriptscriptstyle\rightharpoonup$}} 
\over k}  - \frac{{\mathord{\buildrel{\lower3pt\hbox{$\scriptscriptstyle\rightharpoonup$}} 
\over p} }}{2}} \right)} \phi ^* \left( { - \mathord{\buildrel{\lower3pt\hbox{$\scriptscriptstyle\rightharpoonup$}} 
\over k}  + \frac{{\mathord{\buildrel{\lower3pt\hbox{$\scriptscriptstyle\rightharpoonup$}} 
\over p} }}{2}} \right)d\mathord{\buildrel{\lower3pt\hbox{$\scriptscriptstyle\rightharpoonup$}} 
\over p},
\end{equation}
where $\phi ^*$ is the complex conjugate of $\phi$, $\mathord{\buildrel{\lower3pt\hbox{$\scriptscriptstyle\rightharpoonup$}} 
\over p}$ in the formula is actually the same dimensional as $\mathord{\buildrel{\lower3pt\hbox{$\scriptscriptstyle\rightharpoonup$}} 
\over k}$.
\\
\indent
We take a one-dimensional minimum uncertainty coherent state packet as an example
\begin{equation}
\Phi \left( k \right) = \frac{1}{{\sqrt \alpha  \pi ^{{1 \mathord{\left/
 {\vphantom {1 4}} \right.
 \kern-\nulldelimiterspace} 4}} }}e^{ - \frac{{k^2 }}{{2\alpha ^2 }}}.
\end{equation}
After the first pulse, the momentum space wave function is
\begin{equation}
\Phi \left( k \right) = \frac{1}{{\sqrt {2\alpha } \pi ^{{1 \mathord{\left/
 {\vphantom {1 4}} \right.
 \kern-\nulldelimiterspace} 4}} }}e^{ - \frac{{k^2 }}{{2\alpha ^2 }}}  + \frac{1}{{2\sqrt \alpha  \pi ^{{1 \mathord{\left/
 {\vphantom {1 4}} \right.
 \kern-\nulldelimiterspace} 4}} }}e^{ - \frac{{\left( {k + 2k_0 } \right)^2 }}{{2\alpha ^2 }}} \frac{1}{{2\sqrt \alpha  \pi ^{{1 \mathord{\left/
 {\vphantom {1 4}} \right.
 \kern-\nulldelimiterspace} 4}} }}e^{ - \frac{{\left( {k - 2k_0 } \right)^2 }}{{2\alpha ^2 }}},
\end{equation}
and the Wigner function is
\begin{equation}
\begin{array}{l}
 W\left( {x,k} \right) = e^{ - \frac{{k^2 }}{{\alpha ^2 }}} e^{ - \alpha ^2 x^2 }  + \frac{1}{2}e^{ - \frac{{\left( {k - 2k_0 } \right)^2 }}{{\alpha ^2 }}} e^{ - \alpha ^2 x^2 }  + \frac{1}{2}e^{ - \frac{{\left( {k + 2k_0 } \right)^2 }}{{\alpha ^2 }}} e^{ - \alpha ^2 x^2 }  \\ 
  + \cos \left( {4k_0 x} \right)e^{ - \frac{{k^2 }}{{\alpha ^2 }}} e^{ - \alpha ^2 x^2 }  + \sqrt 2 \cos \left( {2k_0 x} \right)e^{ - \frac{{\left( {k - k_0 } \right)^2 }}{{\alpha ^2 }}} e^{ - \alpha ^2 x^2 }  \\ 
  + \sqrt 2 \cos \left( {2k_0 x} \right)e^{ - \frac{{\left( {k + k_0 } \right)^2 }}{{\alpha ^2 }}} e^{ - \alpha ^2 x^2 }  \\ 
 \end{array}.
\end{equation}
The result shows that after the first square pulse, there are three regions with central momentum $0\hbar k_0$ and $\pm 2\hbar k_0$. In addition, there are three fringe regions associated with the interference term, two of which are located in the middle of $0\hbar k_0$ and $\pm 2\hbar k_0$ momentum regions respectively, and one overlaps with the region of zero momentum. In Fig.4(a), we give an example. The initial state is Gaussian wave packet with temperature $30 nK$. The phase determined by the optical lattice in coordinate space varies periodically with the lattice constant $d = {\lambda  \mathord{\left/
 {\vphantom {\lambda  2}} \right.
 \kern-\nulldelimiterspace} 2}$, where $\lambda  = 532nm$. The fringes in the interference region are vertical after the first pulse, indicating that the phases of the different momentum states are consistent at this time.
\\
\indent
\begin{figure}[h!]
\centering\includegraphics[width=7cm]{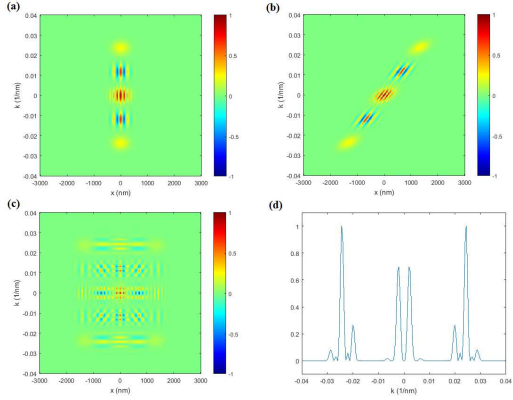}
\caption{The evolution of phase space distribution during the double standing wave pulse beam splitting. (a) The phase space distribution just after the first light pulse. (b)The phase space distribution after free evolution with the dimensionless time $t_2  = 5$. (c) The phase space distribution just after the second light pulse. (d) The momentum distribution after the second light pulse.}
\end{figure}
\\
\indent
When the external potential is zero, the evolution of Wigner function follows
\begin{equation}
W\left( {\mathord{\buildrel{\lower3pt\hbox{$\scriptscriptstyle\rightharpoonup$}} 
\over x} ,\mathord{\buildrel{\lower3pt\hbox{$\scriptscriptstyle\rightharpoonup$}} 
\over k} ,t} \right) = W\left( {\mathord{\buildrel{\lower3pt\hbox{$\scriptscriptstyle\rightharpoonup$}} 
\over x}  - \frac{{\hbar \mathord{\buildrel{\lower3pt\hbox{$\scriptscriptstyle\rightharpoonup$}} 
\over k} }}{m}\delta t,\mathord{\buildrel{\lower3pt\hbox{$\scriptscriptstyle\rightharpoonup$}} 
\over k} ,t_0 } \right),
\end{equation}
here $\delta t = t - t_0$. After free evolution, regions with central momentum of $0\hbar k_0$ and $\pm 2\hbar k_0$ tend to separate and the fringes of the interference region tilt. The fringe inclination in the interference region indicates that different momentum states have different phases. Interestingly, through the free evolution of the Wigner function in phase space, we can clearly see the corresponding relationship between the phase difference and the positions of different momentum states. Here we give an example when the dimensionless time $t_2  = 5$, the three parts with momentum $0\hbar k_0$ and $\pm 2\hbar k_0$ completely separate in space, as shown in Fig. 4(b). The tilt of fringe in the interference regions indicates that the phases of different momentum components due to finite temperature are different after free evolution. Since the fringe inclination is so much, the projection of fringes in the coordinate space becomes smoothing and the periodic variation disappears, that means the coherence between the three parts is hidden. When the second square pulse is applied, the population of component with zero central momentum is simply the sum of zero central momentums coming from three parts under the action of this standing wave pulse. If we consider the population in the zero momentum after the second pulse verse the interval time $\tau _2$ as an interference fringe, then the contrast of the interference fringe reduces to zero after this time. This is consistent with our common understanding that interference does not occur once wave packets do not overlap. Fig. 4(c) shows the phase space distribution after the second pulse. After integrating the coordinates, we got the momentum distribution at this time, as shown in Fig. 4(d). 
\\
\indent
\begin{figure}[h!]
\centering\includegraphics[width=7cm]{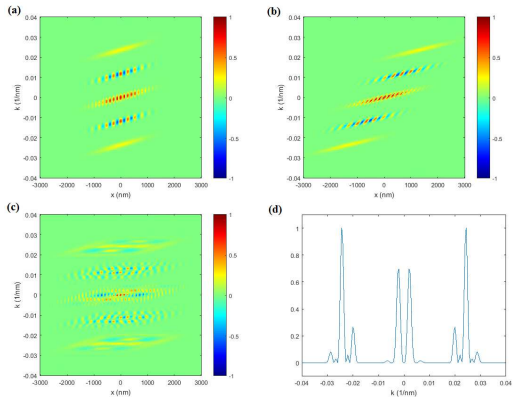}
\caption{The evolution of phase space distribution during the double standing wave pulse beam-splitting when the initial Gaussian wave packet expands before the first pulse. (a) The phase space distribution just after the first light pulse. (b)The phase space distribution after free evolution with the dimensionless time $t_2  = 5$. (c) The phase space distribution just after the second light pulse. (d) The momentum distribution after the second light pulse.}
\end{figure}
\\
\indent
Nevertheless, we want to point out that the overlap of wave packets does not guarantee that the interference fringe contrast will not drop to zero. As an example, we consider the case where the least uncertain Gaussian wave packet diffuses freely for the dimensionless time $t_{{\rm{diff}}}  = 20$ before applying the first square pulse, and other experimental conditions are the same as those in Fig. 4. Just after the first pulse, the phase space distribution is shown in Fig. 5(a). After the same free evolution time, the three wave packets with different momentum are not entirely separated in space, and the inclination of the fringes does not make the projection in the coordinate space disappearing, as shown in Fig. 5(b). However, for any two adjacent groups of fringes, the distance between fringe centers in coordinate positions after free evolution is not equal to the optical lattice constant $d$, so there is a phase shift between the fringes. If we divide and superimpose the projection with the lattice constant $d$, the distribution is still smooth without fringe. Fig. 5(c) shows the phase space distribution after the second pulse, which is different from Fig. 4(c). But the momentum distribution, as shown in Fig. 5(d), is the same as that in Fig. 4(d). As we emphasized earlier, the final momentum distribution depends on the initial momentum distribution and is independent of the spatial distribution. Therefore, whether the wave packet expands before the first pulse does not affect the measurement result. Previous studies have shown that the coherent length of matter waves does not change when the wave packet expands \cite{040}. Our results are consistent with this conclusion.

\section{V. Discussion}

We would like to discuss the defects of existing methods for measuring the temperature of ultra-cold atoms, as well as the unique advantages of our proposed approach.
\\
\indent
The commonly used TOF method requires a free expansion time of hundreds of milliseconds to measure a temperature on the order of $100 pK$. For ground systems, this is simply not possible due to gravity. Long-term TOF can be achieved in microgravity environments, but many experimental measurements are still required for fitting \cite{017}. In addition, if there is a weak interaction between atoms, such as BECs, the interaction potential energy convert to kinetic energy continuously during a long period of expansion, so the TOF method is not actually suitable for measuring the change in the momentum width over time after being released from the trap.
\\
\indent
Bragg Spectroscopy is another commonly used method to obtain the distribution of atomic velocities \cite{019,020}. To measure a narrow velocity distribution, the frequency stability of Bragg beams with tens of Hz is necessary. For example, when the temperature of ${}^{87}{\rm{Rb}}$ atomic gas is about $100 pK$, the corresponding velocity width is about $0.2 mm/s$. If we set the resolution as $0.02 mm/s$, the frequency difference between two counter-propagating Bragg beams would correspond to about $50 Hz$. The simplest way to create such Bragg beam splitters is to employ two acousto-optic modulators with phase-stable radio frequency drives. However, the laser system of the ultra-cold atomic experiment device on the Chinese space Station separates from the experiment chamber \cite{041}, and optical fibers connected in the middle, which will bring additional phase noise due to the influence of the environment \cite{042}. Therefore, a low noise phase-locked loop (PLL) with fiber output sampling is required. Due to resource constraints, the current system is not equipped with such a PLL. Moreover, a series of experiments are required to obtain the Bragg spectrum. If a single continuous scan is used, the continuous scanning time will also reach about $100 ms$. The optical depth of the atomic cloud diffracted by such long time scanning is very low and the signal-to-noise ratio is poor.
\\
\indent
The atom-optics knife-edge technique is more suitable for measuring the momentum widths of atoms in waveguides \cite{021}. Due to the need to generate a potential barrier in the order of micron width, optical systems with large numerical apertures are required, which is difficult to achieve for some experimental systems due to the geometrical size restrictions. This method also requires a series of experiments and then fitting the results.
\\
\indent
We also looked at other measurement methods, such as detection of the change of atomic densities in a bucket region and the spin gradient thermometry. The former can do nearly nondestructive temperature measurements of cold atomic ensembles, but the measurement range is hard to extend below the order of nanokelvin \cite{043}. The latter in principle allows measurement of temperatures down to $50 pK$, but can only be used under certain conditions \cite{044}. In recent years, quantum thermometry has aroused great interest in researchers, which also has the potential to measure sub-nK temperature \cite{045,046,047}. However, no experiment so far has demonstrated its practical feasibility in ultracold atoms.
\\
\indent
Compared with the above methods, our approach is simple in structure and timing control. For many cold atom experimental setups, one-dimensional optical lattices already exist, so the measurements can be made with only the appropriate timing. Since we measure the ratio of the population in each momentum state, the fluctuation of atom number only has very slight effect.
\\
\indent
In dealing with the case of BECs, we ignore the effects of atomic interactions. The interaction between atoms in evolution provides an additional phase. We use the chemical potential of $100 pK$ as an example to estimate the additional phase. Assuming the laser wavelength of the optical lattice is $532 nm$, the free flight of about 10ms after the double pulse process can completely separate the center zero momentum part from the $\pm 2\hbar k_0$ momentum part in space. The maximum value of phase imprinting after all the evolution time is $\frac{{k_B T\tau }}{\hbar } \approx 0.157 \approx 0.05\pi$ and the actual inhomogeneous phase is less than this value. This shows it is a reasonable approximation to ignore the interactions between atoms.
\\
\indent
We also estimate the effect of the optical lattice vibration during the actual measurement period \cite{048}. In the space station, the acceleration can be reduced to no more than a magnitude of $10^{ - 4} {m \mathord{\left/
 {\vphantom {m {s^2 }}} \right.
 \kern-\nulldelimiterspace} {s^2 }}$ (RMS) at full frequency after vibration isolation \cite{049}. In the double-pulse beam-splitting period, the position of the optical lattice drifts in the order of nanometer, which is much smaller than the lattice constant d of the optical lattice. Thus, this method is valid to measure the momentum distribution of ultra-cold atoms approximately $100 pK$, even after taking into account the actual vibrations on the space station.
\\
\indent
In the experimental implementation, the shorter wavelength of optical lattice laser is more advantageous, which will bring three benefits: 1) At the same temperature, the shorter the wavelength of the optical lattice laser, the smaller the random momentum of the atom relative to the recoil momentum of the optical lattice. Therefore, the standing wave pulse can be better approximated as a Hadamard operation. 2) The optical lattice transfers more recoil momentum to the atoms, so that the complete separation of the wave packet takes less time and the interaction between the atoms causes less impact. 3) The required interval $\tau _2$ is shorter and the measurement results are less sensitive to vibration.
\\
\indent
Recently, we did a simulation experiment on an optical system. We also measure the momentum width of BEC in the optical trap by the above method. Preliminary experimental results demonstrate that the scheme is viable and can be used to measure ultra-narrow momentum distribution.

\section{VI. Conclusion}

We have proposed a technique for characterizing ultralow-velocity width of atomic gases, which requires only a one-dimensional optical lattice and simple timing control. Since what we need to measure is the ratio of population of each momentum state, the effect of atomic number fluctuation is not significant. And, in principle, a single shot can obtain the outcome. The scheme has the ability to measure ultra-cold atomic samples at temperatures even lower than $100 pK$, providing temperature calibration for cold atomic physics experiments on the space station. 
\bigskip
\bigskip
\bigskip
\bigskip

\section{Acknowledgments}
\begin{acknowledgments}
We thank Prof. Changhe Zhou for helpful discussions and Ms. Xuemei Jia for her help. This work was supported by the National Key Research and Development Program of China (No. 2021YFA0718303) and the National Natural Science Foundation of China under Grant No.11674338, No.U1730126 and No.11547024, and funds provided by Technology and Engineering Center for Space Utilization, Chinese Academy of Sciences.
\end{acknowledgments}

\begin{thebibliography}{0}%
\makeatletter
\providecommand \@ifxundefined [1]{%
 \@ifx{#1\undefined}
}%
\providecommand \@ifnum [1]{%
 \ifnum #1\expandafter \@firstoftwo
 \else \expandafter \@secondoftwo
 \fi
}%
\providecommand \@ifx [1]{%
 \ifx #1\expandafter \@firstoftwo
 \else \expandafter \@secondoftwo
 \fi
}%
\providecommand \natexlab [1]{#1}%
\providecommand \enquote  [1]{``#1''}%
\providecommand \bibnamefont  [1]{#1}%
\providecommand \bibfnamefont [1]{#1}%
\providecommand \citenamefont [1]{#1}%
\providecommand \href@noop [0]{\@secondoftwo}%
\providecommand \href [0]{\begingroup \@sanitize@url \@href}%
\providecommand \@href[1]{\@@startlink{#1}\@@href}%
\providecommand \@@href[1]{\endgroup#1\@@endlink}%
\providecommand \@sanitize@url [0]{\catcode `\\12\catcode `\$12\catcode
  `\&12\catcode `\#12\catcode `\^12\catcode `\_12\catcode `\%12\relax}%
\providecommand \@@startlink[1]{}%
\providecommand \@@endlink[0]{}%
\providecommand \url  [0]{\begingroup\@sanitize@url \@url }%
\providecommand \@url [1]{\endgroup\@href {#1}{\urlprefix }}%
\providecommand \urlprefix  [0]{URL }%
\providecommand \Eprint [0]{\href }%
\providecommand \doibase [0]{https://doi.org/}%
\providecommand \selectlanguage [0]{\@gobble}%
\providecommand \bibinfo  [0]{\@secondoftwo}%
\providecommand \bibfield  [0]{\@secondoftwo}%
\providecommand \translation [1]{[#1]}%
\providecommand \BibitemOpen [0]{}%
\providecommand \bibitemStop [0]{}%
\providecommand \bibitemNoStop [0]{.\EOS\space}%
\providecommand \EOS [0]{\spacefactor3000\relax}%
\providecommand \BibitemShut  [1]{\csname bibitem#1\endcsname}%
\let\auto@bib@innerbib\@empty
\end{thebibliography}%


\begin{thebibliography}{000}
\bibitem{001} S. Nimmrichter and K. Hornberger, Macroscopicity of Mechanical Quantum Superposition States, Phys. Rev. Lett. 110, 160403 (2013).
\bibitem{002} A. Bassi, K. Lochan, S. Satin, T. Singh, and H. Ulbricht, Models of wave-function collapse, underlying theories, and experimental tests, Rev. Mod. Phys. 85, 471 (2013).
\bibitem{003} M. J. H. Ku,  A. T. Sommer, L. W. Cheuk, and M. W. Zwierlein, Revealing the superfluid lambda transition in the universal thermodynamics of a unitary Fermi gas, Science 335, 563 (2012).
\bibitem{004} J. Léonard, A. Morales, P. Zupancic, T. Esslinger, and T. Donner, Supersolid formation in a quantum gas breaking a continuous translational symmetry, Nature (London) 543, 87 (2017).
\bibitem{005} J. Li, J. Lee, W. Huang, S. Burchesky, B. Shteynas, F. Ç. Top, A. O. Jamison, and W. Ketterle, A stripe phase with supersolid properties in spin–orbit-coupled Bose–Einstein condensates, Nature (London) 543, 91 (2017).
\bibitem{006} L. X. Niu, S. J. Jin, X. Z. Chen, X. P. Li, and X. J. Zhou, Observation of a dynamical sliding phase superfluid with P-band Bosons, Phys. Rev. Lett. 121, 265301 (2018).
\bibitem{007} V. Efimov, Energy levels arising from resonant two-body forces in a three-body system, Phys. Lett. 33B, 563 (1970).
\bibitem{008} T. Kraemer, M. Mark, P. Waldburger, J.G. Danzl, C. Chin, B. Engeser, A.D. Lange, K. Pilch, A. Jaakkola, H.-C. Nägerl, and R. Grimm, Evidence for Efimov quantum states in an ultracold gas of caesium atoms, Nature (London) 440, 315 (2006).
\bibitem{009} J. P. D’Incao, H. Suno, and B. D. Esry, Limits on Universality in ultracold three-Boson recombination, Phys. Rev. Lett. 93, 123201 (2004).
\bibitem{010} A. D. Cronin, D. E. Pritchard, and J. Schmiedmayer, Optics and interferometry with atoms and molecules, Rev. Mod. Phys. 81, 1051 (2009).
\bibitem{011} J.M. Hogan, D.M. S. Johnson, S. Dickerson, T. Kovachy, A. Sugarbaker, S.-w. Chiow, P.W. Graham,M. A. Kasevich, B. Saif, S. Rajendran, P. Bouyer, B. D. Seery, L. Feinberg, and R. Keski-Kuha, An atomic gravitational wave interferometric sensor in low earth orbit (AGIS-LEO), Gen. Relativ. Gravit. 43, 1953 (2011).
\bibitem{012} A. E. Leanhardt, T. A. Pasquini, M. Saba, A. Schirotzek, Y. Shin, D. Kielpinski, D. E. Pritchard, and W. Ketterle, Cooling Bose-Einstein condensates below 500 picokelvin, Science 301, 1513 (2003).
\bibitem{013} T. Luan, H. P. Yao, L. Wang, C. Li, S. F.Yang, X. Z. Chen, and Z. Y. Ma, Two-stage crossed beam cooling with 6Li and 133Cs atoms in microgravity, Opt. Express 23, 011378 (2015).
\bibitem{014} S. Chu, J. E. Bjorkholm, A. Ashkin, J. P. Gordon, and L. W. Hollberg, Proposal for optically cooling atoms to temperatures of the order of 10-6K, Opt. Lett. 11, 73 (1986).
\bibitem{015} H. Ammann and N. Christensen, Delta kick cooling: A new method for cooling atoms, Phys. Rev. Lett.78, 2088 (1997).
\bibitem{016} T. Kovachy, J. M. Hogan, A. Sugarbaker, S. M. Dickerson, C. A. Donnelly, C. Overstreet, and M. A. Kasevich, Matter wave lensing to picokelvin temperatures, Phys. Rev. Lett. 114, 143004 (2015).
\bibitem{017} D. C. Aveline, J. R. Williams, E. R. Elliott, C. Dutenhoffer, J. R. Kellogg, J. M. Kohel, N. E. Lay, K. Oudrhiri, R. F. Shotwell, N. Yu, and R. J. Thompson, Observation of Bose–Einstein condensates in an Earth-orbiting research lab, Nature 582,193 (2020).
\bibitem{018} C. Deppner, W. Herr, M. Cornelius, P. Stromberger, T. Sternke, C. Grzeschik, A. Grote, J. Rudolph, S. Herrmann, M. Krutzik, A. Wenzlawski, R. Corgier, E. Charron , D. Guéry-Odelin, N. Gaaloul, C. Lämmerzahl, A. Peters, P. Windpassinger, and E. M. Rasel, Collective-mode enhanced matter-wave optics, Phys. Rev. Lett. 127, 100401 (2021).
\bibitem{019} J.-Y. Courtois, G. Grynberg, B. Lounis, and P. Verkerk, Recoil-induced resonances in cesium: An atomic analog to the free-electron laser, Phys. Rev. Lett. 72, 3017(1994).
\bibitem{020} J. Stenger, S. Inouye, A. P. Chikkatur, D. M. Stamper-Kurn, D. E. Pritchard, and W. Ketterle, Bragg Spectroscopy of a Bose-Einstein Condensate, Phys. Rev. Lett. 82, 4569 (1999).
\bibitem{021} R. Ramos, D. Spierings, S. Potnis, and A. M. Steinberg, Atom-optics knife edge: Measuring narrow momentum distributions, Phys. Rev. A 98, 023611 (2018).
\bibitem{022} S. S. Szigeti, J. E. Debs, J. J. Hope, N. P. Robins and J. D. Close, Why momentum width matters for atom interferometry with Bragg pulses, New J. Phys. 14, 023009 (2012).
\bibitem{023} G. D. McDonald, C. C. N. Kuhn, S. Bennetts, J. E. Debs, K. S. Hardman, M. Johnsson, J. D. Close, and N. P. Robins, 80$\hbar k$ momentum separation with Bloch oscillations in an optically guided atom interferometer, Phys. Rev. A 88, 053620 (2013).
\bibitem{024} M. Carey, J. Saywell, D. Elcock, M. Belal, and T. Freegarde, Velocimetry of cold atoms by matter-wave interferometry, Phys. Rev. A 99, 023631 (2019).
\bibitem{025} L. P. Parazzoli, A.M. Hankin and G.W. Biedermann, Observation of free-space single-atom matter wave interference, Phys. Rev. Lett. 109, 230401 (2012).
\bibitem{026} S. Wu, Y.-J. Wang, Q. Diot, and M. Prentiss, Splitting matter waves using an optimized standing-wave light-pulse sequence, Phys. Rev. A 71, 043602 (2005).
\bibitem{027} Y.-J. Wang, D. Z. Anderson, V. M. Bright, E. A. Cornell, Q. Diot, T. Kishimoto, M. Prentiss, R.A. Saravanan, S. R. Segal, and S. Wu, Atom Michelson interferometer on a chip using a Bose-Einstein condensate, Phys. Rev. Lett. 94, 090405 (2005). 
\bibitem{028} W. D. Montgomery, Self-imaging objects of infinite aperture, J. Opt. Soc. Am. 57, 772 (1967).
\bibitem{029} J. M. Wen, Y. Zhang, and M. Xiao, The Talbot effect: recent advances in classical optics, nonlinear optics, and quantum optics, Advances in Optics and Photonics 5, 83 (2013).
\bibitem{030} H. K. Andersen, Bose-Einstein condensates in optical lattices, Ph.D. thesis, University of Aarhus, 2008. 
\bibitem{031} J. H. Denschlag, J. E. Simsarian, H. Häffner, C. McKenzie, A. Browaeys, D. Cho, K. Helmerson, S. L. Rolston and W. D. Phillips, A Bose–Einstein condensate in an optical lattice, J. Phys. B: At. Mol. Opt. Phys. 35, 3095 (2002).
\bibitem{032} M. Gondran and A. Gondran, Numerical simulation of the double slit interference with ultracold atoms, Am. J. Phys. 73, 507 (2005).
\bibitem{033} C. J. Pethick and H. Smith, Bose-Einstein Condensation in Dilute Gases  (Cambridge University Press, Cambridge, 2002).
\bibitem{034} Y. Castin and R. Dum, Bose-Einstein condensates in time dependent traps, Phys. Rev. Lett. 77, 5315 (1996).
\bibitem{035} Yu. Kagan, E. L. Surkov, and G. V. Shlyapnikov, Evolution of a Bose-condensed gas under variations of the confining potential, Phys. Rev. A 54, R1753 (1996).
\bibitem{036} C. Y. Wong, Explicit solution of the time evolution of the Wigner function, J. Opt. B: Quantum Semiclass. Opt. 5, S420 (2003).
\bibitem{037} C. Kurtsiefer, T. Pfau and J. Mlynek, Measurement of the Wigner function of an ensemble of helium atoms, Nature (London) 386, 150 (1997). 
\bibitem{038} S. Y. Zhou, J. Chabé, R. Salem, T. David, D. Groswasser, M. Keil, Y. Japha, and R. Folman, Phase space tomography of cold-atom dynamics in a weakly corrugated potential, Phys. Rev. A 90, 033620 (2014).
\bibitem{039} A. Fannjiang, J. Shi, and G. Papanicolaou, High frequency behavior of the focusing nonlinear Schrödinger equation with random inhomogeneities, SIAM J. Appl. Math., 63, 1328 (2003).
\bibitem{040} J. R. Kellogg, N. Yu, J. M. Kohel, R. J. Thompson, D. C. Aveline and L. Maleki, Longitudinal coherence in cold atom interferometry, J. Mod. Opt. 54, 2533 (2007).
\bibitem{041} B.Wang, W. Xiong, S. Y. Zhou, D. J. Chen, Q. Z. Qu, T. Li, L. Li, T. Q. Song, W. B. Chen, X. Z. Chen, and L. Liu, Design of an ultra-cold atom setup in microgravity, in The 26th International Conference on Atomic Physics, (Barcelona, 2018).
\bibitem{042} L. S. Ma, P. Jungner, J. Ye, Hall, J. L. Hall, Delivering the same optical frequency at two places: accurate cancellation of phase noise introduced by an optical fiber or other time-varying path, Opt. Lett. 19, 1777 (1994).
\bibitem{043} X. Wang, Y. Sun, H.-D. Cheng, J.-Y. Wan, Y.-L. Meng, L. Xiao, and L. Liu, Nearly nondestructive thermometry of labeled cold atoms and application to isotropic laser cooling, Phys. Rev. Applied 14, 024030 (2020).
\bibitem{044} D. M. Weld, P. Medley, H. Miyake, D. Hucul, D. E. Pritchard, and W. Ketterle, Spin gradient thermometry for ultracold atoms in optical lattices, Phys. Rev. Lett. 103, 245301 (2009).
\bibitem{045} S. Jevtic, D. Newman, T. Rudolph, and T. M. Stace, Single-qubit thermometry, Phys. Rev. A 91, 012331 (2015).
\bibitem{046} M. Mehboudi, A. Lampo, C. Charalambous, L. A. Correa, M. A. García-March, and M. Lewenstein, Using polarons for Sub-nK quantum nondemolition thermometry in a Bose-Einstein condensate, Phys. Rev. Lett. 122, 030403 (2019).
\bibitem{047} M. T. Mitchison, T. Fogarty, G. Guarnieri, S. Campbell, T. Busch, and J. Goold, In situ thermometry of a cold Fermi gas via dephasing impurities, Phys. Rev. Lett. 125, 080402 (2020).
\bibitem{048} M. Barmatz, Inseob Hahn, J. A. Lipa, and R. V. Duncan, Critical phenomena in microgravity: Past, present, and future, Rev. Mod. Phys. 79, 1 (2007).
\bibitem{049} W. B. Dong, W. X. Duan, W. Liu and Y. K. Zhang, Microgravity disturbance analysis on Chinese space laboratory, NPJ Microgravity 5,18 (2019).

\end{thebibliography}

\end{document}